\documentclass[prl,a4paper,nofootinbib,
twocolumn,
]{revtex4-1}
\usepackage{graphicx}
\usepackage{amsfonts}
\usepackage{amssymb}
\usepackage{slashed} 
\usepackage{color}
\usepackage{hyperref}
\def\eq#1{{Eq.~(\ref{#1})}}
\def\eqs#1#2{{Eqs.~(\ref{#1})--(\ref{#2})}}

\def\di{\mbox{d}}

\definecolor{oucrimsonred}{rgb}{0.6, 0.0, 0.0}
\definecolor{persianblue}{rgb}{0.11, 0.22, 0.73}
\definecolor{forestgreen}{rgb}{0.13,0.35,0.13}
 \hypersetup{colorlinks, citecolor=oucrimsonred, linkcolor=persianblue, urlcolor=oucrimsonred}
\def\hhref#1{\href{http://arxiv.org/abs/#1}{#1}} 
\newcommand{\be}{\begin{equation}}
\newcommand{\ee}{\end{equation}}
\newcommand{\bea}{\begin{eqnarray}}
\newcommand{\eea}{\end{eqnarray}}
\newcommand{\nn}{\nonumber}

\newcommand{\com}[1]{}
\newcommand{\gsim}{\lower.7ex\hbox{$\;\stackrel{\textstyle>}{\sim}\;$}}
\newcommand{\lsim}{\lower.7ex\hbox{$\;\stackrel{\textstyle<}{\sim}\;$}}

\newcommand{\BR}{\mbox{BR\,}}

\begin{document}
\title[]{ Searching for the dark sector in two-body anti-muon decay\\
with the polarization of monochromatic positrons 
}
\date{December 2, 2020}
\author{M.\ Fabbrichesi$^{\dag}$}
\author{E.\ Gabrielli$^{\ddag\dag *}$}
\affiliation{$^{\dag}$INFN, Sezione di Trieste, Via  Valerio 2, 34127 Trieste, Italy }
\affiliation{$^{\ddag}$Physics Department, University of Trieste}
\affiliation{$^{*}$NICPB, R\"avala 10, Tallinn 10143, Estonia }
\begin{abstract}
\noindent  The $\mu^+ \to e^+ X$ decay, where $X$ is a dark sector boson, provides one of the strongest 
available bounds on the scale of dark sector interactions. The $X$  boson  can be an axion or a dark photon. We show that the concurrent determination of the anti-muon and positron  polarizations makes possible to distinguish with a confidence level of 99\% between the two dark sector portals with as few as 6 observed events in the case of the massless dark photon. Instead, the  massive  spin-1, dimension 4 dark portal  cannot be distinguished from the axion-like case. We also discuss the possibility that the $X$ boson be a massive spin-2  particle.
 
\end{abstract}

\maketitle 
\textit{Motivations.---}
The experimental search for the two-body decay $\mu^+\to e^+ X$ can provide a direct window on the hypothetical dark sector~\cite{dark_sector}, at least as long as this sector   contains flavor changing interactions. New experiments are now under way to reach  branching rates (BR) of the order of $O(10^{-8})$~\cite{Perrevoort:2018ttp} at Mu3e~\cite{Blondel:2013ia} and $O(10^{-7})$~\cite{Calibbi:2020jvd} at  MEG-II~\cite{Baldini:2018nnn}.  Given the high sensitiveness involved, they might even turn the limit into a discovery. If this turns out to be the case, it will be imperative to identify what kind of particle is the dark sector $X$ boson. The two main candidates are an axion-like particle (ALP) and a dark photon.
 As we show below, these two particles give rise to  decay widths that are rather similar (identical in the massless case) if  the positron polarization is not measured. On the contrary, the measurement of the positron polarization provides a very efficient way to determine the nature of the $X$ boson and decide what kind of portal to the dark sector has been discovered.

\vskip2em
\textit{The differential probabilities.---}  The  decay of anti-muons into positrons  in the Standard Model (SM) represents the background of any search for the decay $\mu^+ \to e^+ X$. It 
 is a three-body decay with the corresponding two neutrinos and the positron in the final states. The differential probability for emitting a positron, with a reduced energy $x_e=2 m_\mu E_e/ (m_\mu^2 + m_e^2) \simeq 2 E_e/m_\mu$
 and a momentum with an angle  $\theta$ to the muon spin axis,  is given by~\cite{Kuno:1999jp,Gorringe:2015cma}
\bea
\frac{\di \Gamma (\mu^+\to e^+ \bar \nu_\mu \nu_e)  }{\di x_e \, \di \cos \theta}&  =&  \frac{G_F^2 m_\mu^5}{192 \pi^3} x_e^2 \frac{\left( 1 + \lambda_e   \right)}{2}   \label{gamma}\\
&\!\!\!\!\! \!\!\!\!\!  \times &  \!\!\!\!\! \Big[ 3 - 2 x _e +P_\mu (2 x _e- 1) \cos \theta \Big] \nn
\label{SM}
\eea
where the mass of the positron is neglected; $P_\mu=|\vec{P_\mu}|$ and $\lambda_e$ are the polarization vector of the anti-muon and the helicity of the positron, respectively.   

\begin{figure}[t!]
\begin{center}
\includegraphics[width=3in]{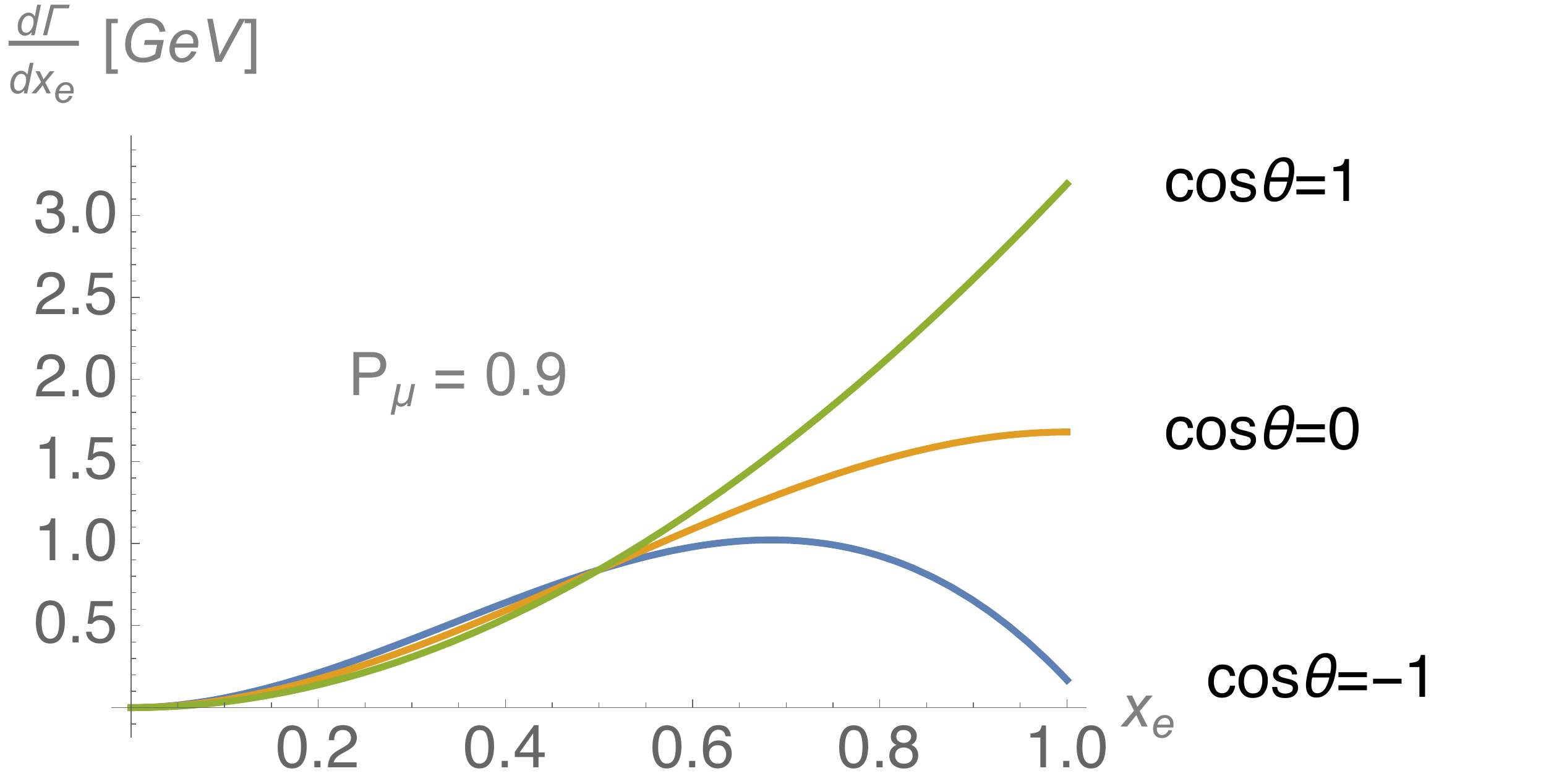}
\caption{\small
  Dependence of the  differential width for the \hbox{$\mu^+\to e^+ \bar \nu_\mu \nu_e$} decay in the SM on the reduced energy $x_e$ for an averaged muon polarization of 90\%. Three choices for $\cos \theta$ are shown.
\label{sm} 
}
\end{center}
\end{figure}

For polarized anti-muons, the SM differential width in \eq{gamma}  is suppressed for certain values of  $\cos \theta$ at the spectrum endpoint (see Fig.~\ref{sm}). This feature has been exploited by some of the experimental searches. For positrons with helicity $\lambda_e = -1$ the width in \eq{gamma}  vanishes (we neglect terms proportional to the positron mass and transverse polarizations) because of the chiral structure of the SM charged currents. This feature too can be used to control the SM background contribution to the events.

 Axion-like particles~\cite{Ringwald:2014vqa,Graham:2015ouw}  are pseudo-scalar bosons that can mediate the two-body anti-muon decay into positron $\mu^+\to e^+X$ with an interaction Lagrangian given by 
\be
{\cal L}_{\rm ALP}=\frac{c_{\mu e}^L}{f_a}  \, \bar \mu_R  \gamma^\mu e_R \, \partial_\mu a +
 \frac{c_{\mu e}^R}{f_a}  \, \bar \mu_L  \gamma^\mu e_L\, \partial_\mu a \label{L-ALP}\, + {\rm H.c.}\, ,
\ee
where $L$ and $R$ indicate the respective chiral spinors.
We do not consider the vector-like structure and only retain the chiral couplings (see \cite{Calibbi:2020jvd} for a recent discussion of all possible coupling structures).

The dark photon~\cite{Fabbrichesi:2020wbt} is a vector boson, massless or massive, that provides the same two-body decay with a gauge-invariant interaction Lagrangian given by
\be
{\cal L}_{\rm DP}=\frac{d_{\mu e}^L}{2\Lambda} \, \bar \mu_R \sigma_{\mu\nu} e_L F^{\mu \nu} +
\frac{d_{\mu e}^R}{2\Lambda} \, \bar \mu_L \sigma_{\mu\nu} e_R F^{\mu \nu}  \label{L-DP}\, +{\rm H.c.}\, ,
\ee
where $\sigma_{\mu\nu} =( i/2) [ \gamma_\mu, \gamma_\nu ]$ and $F^{\mu \nu} $ is the dark photon fields strength.

The dipole interaction in the form of \eq{L-DP} is the  leading interaction between the SM states and the massless dark photon for flavor changing neutral current (FCNC) processes, which is also the same for the massive dark-photon. In the massive case, because  the $U(1)_D$ gauge symmetry is broken, tree-level FCNC interactions mediated by dimension 4 operators can arise, namely
\be
   {\cal L}^{\rm tree}_{\rm MDP}=\bar{d}_{\mu e}^L \, \bar \mu_L \gamma_{\mu}  e_L A^{\mu} +
                  \bar{d}_{\mu e}^R \, \bar \mu_R \gamma_{\mu}  e_L A^{\mu}   
\label{L-MDP}\, +{\rm H.c.}\, ,
\ee
where $A^{\mu}$ is the massive dark-photon field. 

For the FCNC dipole interaction in \eq{L-DP}, the massless limit $m_X\to 0$ of the corresponding massive dark-photon amplitude is smooth for the process $\mu\to e X$ and going into the  amplitude for the massless case. The only mass discontinuity between massive and massless scenarios arises in the flavor conserving processes for which there is a direct interaction to the SM current for the massive case whereas the dipole operator remains the only interaction to SM fermions in the massless scenario~\cite{Dobrescu:2004wz}.
On the other hand, for the case of \eq{L-MDP}, since the longitudinal component of massive dark-photon does not decouple in $m_X\to 0$, there is no corresponding massless limit.

The interactions in \eq{L-ALP} and \eq{L-DP} or \eq{L-MDP} provide a two-body channel for the muon to decay into and a characteristic signal of fixed-energy positrons  at the endpoint of the SM spectrum (see Fig.~\ref{sm}) at $x_e=1$ for massless $X$ boson. If the dark photon or the ALP are massive the contribution is shifted to values $x_e<1$. 
The positron momentum $p_e=|{\bf \vec{p}_e}|$ and energy $E_e$ are given by
\bea
p_e &=&\frac{1}{2 m_\mu}  \sqrt{\Big[(m_\mu + m_e)^2 - m_X^2 \Big]\Big[(m_\mu -
    m_e)^2 - m_X^2 \Big] }\, , \nn \\
E_e &=&\frac{1}{2 m_\mu}\left(m_{\mu}^2+m_e^2-m_X^2\right)\, .
\eea
where $m_X$ stands for the mass of the  dark boson $X$. We assume that the $X$ boson is stable or decays outside the detector (see \cite{Heeck:2017xmg} for a discussion  that also includes the case in which it decays inside the detector).

The corresponding differential probabilities are
\begin{widetext}
\bea
\left. \frac{\di \Gamma}{\di\cos \theta} \right|_{\text{DP}} &=& \frac{m_\mu^3}{\Lambda^2} \frac{ (1 - r)^2}{ 64 \pi}  
\Big\{ |d_{\mu e}^R|^2 (1- \lambda_e)  \left[ 1+  P_\mu   \cos \theta  + \frac{r}{2} ( 1 - P_\mu   \cos \theta) \right]  \Big.\nn  \\
& & +  \Big. |d_{\mu e}^L|^2(1+ \lambda_e)  \left[ 1- P_\mu   \cos \theta  + \frac{r}{2} ( 1 + P_\mu   \cos \theta) \right]\Big\}   \label{widthDP} \\
\left. \frac{\di \Gamma}{\di\cos \theta}\right| _{\text{ALP}} &=& \frac{m_\mu^3}{f_a^2} \frac{(1 - r)^2}{ 128 \pi} 
\Big\{ |c_{\mu e}^L|^2 (1-\lambda_e) (1- P_\mu   \cos \theta ) + |c_{\mu e}^R|^2 (1+ \lambda_e) (1+  P_\mu   \cos \theta )\Big\} \, ,
\label{widthALP}
\nonumber \\
\\
\left. \frac{\di \Gamma}{\di\cos \theta} \right|^{\rm tree}_{\text{MDP}} &=& \frac{m_\mu^3}{m_X^2} \frac{ (1 - r)^2}{ 64 \pi}  
\Big\{ |\bar{d}_{\mu e}^R|^2 (1- \lambda_e)  \left[ 1-  P_\mu   \cos \theta  + 2r ( 1 + P_\mu   \cos \theta) \right]  \Big.\nn  \\
& & +  \Big. |\bar{d}_{\mu e}^L|^2(1+ \lambda_e)  \left[ 1+ P_\mu   \cos \theta  + 2r ( 1 - P_\mu   \cos \theta) \right]\Big\}   \label{widthDPM}
\eea
\end{widetext}
where $r=m_X^2/m_\mu^2$ and we neglect the mass of the positron. While the width for  the massive dark photon mediated by dipole interaction contains a  term proportional to the dark-photon mass due to the contribution of the longitudinal component (which behaves as a scalar field) the scalar case has no mass corrections and the massive and massless scalar cases have the same angular distribution.
As we can see, in the case of the FCNC tree-level interaction in \eq{L-MDP}, the massive dark-photon behaves as the ALP in the $m_X\ll m_{\mu}$ limit, as far as polarizations and angular distributions are concerned, provided a redefinition
of the couplings $c^L_{\mu e} \leftrightarrow \bar{d}^R_{\mu e}, c^R_{\mu e}  \leftrightarrow  \bar{d}^L_{\mu e}$ is adopted. Indeed, in this case only the longitudinal polarization of massive dark-photon field dominates, with an effective coupling inversely proportional to the dark-photon mass. Then, since we will restrict to the small mass scenario $m_X\ll m_{\mu}$, we will not consider the analysis of the massive dark-photon contribution mediated by tree-level FCNC, this case being included into the ALP one. That said, one should bear in mind that the massive dark photon coupled through the currents in \eq{L-MDP} is indistinguishable from the ALP.

For the massless case, the widths in \eq{widthDP} and \eq{widthALP} become the same  if the positron helicity is summed over, with the left-handed interaction exchanged with the right-handed in going from ALP to dark photons. For a massive dark photon, making no assumptions on the coefficients, the two widths are different and the difference is controlled by the value of $m_X$. This difference is too small to be very efficient in telling the spin of the dark sector boson. Even for sizable values of the mass $m_X$, the discriminating power is not optimal---as we show in the following.
To distinguish between the spin 0 and 1 option in an efficient manner, we need to measure the polarization of the positrons. 
Since  we do not expect to have many events to use in the analysis---given the rarity of these branching ratios---it is important to be able to discriminate the two spin hypotheses with as few events as possible.

The simultaneous measurement of the positron polarization and angular distribution of the photon momentum always allows to disentangle the spin-0 ALP from the massless spin-1 dark-photon, regardless of the relative weights of left-handed and right-handed couplings appearing in the interaction vertices. Indeed, the measurement of the positron polarization uniquely selects the specific chiral couplings in the ALP and dark-photon in a correlated way, thus selecting the corresponding angular distributions in the spin-0 ALP and spin-1 sectors. This correlation is a consequence of  angular momentum conservation in the $\mu \to e X$ decay.

The simultaneous measurement of the positron polarization and angular distribution of photon  is also useful for controlling  the SM background. It makes possible to select  the suppressed SM backgrounds by means of  the choice of  positron polarization $\lambda_e$. For example, the request of having $\lambda_e=-1$ in the final positron states can strongly suppress the leading SM contribution, which is mainly associated to the $\lambda_e=1$ polarization, as it is evident from  Eq.(\ref{SM}), where only the leading contribution in the $m_e\to 0$ limit is retained.

\vskip0.5cm
\textit{Experimental results.---} Limits  for the decay $\mu^+ \to e^+ X$ are currently based on two experiments, both  based at the TRIUMF laboratory. They use  different experimental strategies.

The older one~\cite{Jodidio:1986mz} exploits the reduction in the SM rate for polarized anti-muons for angles  close to $\cos \theta_e = -1$ (see Fig.\ref{sm}) in the region $x_e=1$. Anti-muons are almost fully polarized, with $P_\mu =0.9$. The SM background is very suppressed at the endpoint of the positron.
For a massless boson $X$, this is also the region of the monochromatic contribution.
They find
\be
\BR (\mu^+ \to e^+ X)  \leq 2.6 \times 10^{-6}\ \quad 90\% \;\text{CL} \label{jodidio0}
\ee
for an assumed familon particle~\cite{Anselm:1985bp} that has a vector-like coupling. The limit  in \eq{jodidio0} has been re-interpreted~\cite{Calibbi:2020jvd} in terms of chiral ALP to give
\be 
\BR (\mu^+ \to e^+ X)  \leq 2.5 \times 10^{-6} \quad 90\%\; \text{CL} \label{jodidio}
\ee
for the right-handed chirality. The re-interpretation of the corresponding limit for the left-handed chirality is much weaker~\cite{Calibbi:2020jvd}. 

The bound on the branching ratio in \eq{jodidio}
implies a very strong limit on the scale of both the right-handed ALP  and the dark photon, namely
\be
\left|\frac{ f_a}{c_{\mu e}^{R}}\right| \quad \text{and} \quad\left| \frac{\Lambda}{\sqrt{2} \,d_{\mu e}^{R}}\right|   \geq 3.9 \times 10^{6} \; \mbox{TeV} \, ,
\label{boundR}
\ee
in agreement with the analysis in \cite{Calibbi:2020jvd} as far as the massless ALP is concerned.
For massive ALP or dark photons, the  limit in \eq{jodidio} can be  applied as long as $m_X < 10$ MeV.

The more recent  experiment  by the TWIST collaboration~\cite{Bayes:2014lxz} does not exploit the polarization of the anti-muons.  The SM background  is larger and they  use a positron endpoint spectrum calibration; they provide a limit for 
left-handed chiral ALP 
\be 
\BR  (\mu^+ \to e^+ X) \leq 5.8 \times 10^{-5} \label{twist}\quad 90\% \;\text{CL}
\ee
as well as for the right-handed one:
\be 
\BR  (\mu^+ \to e^+ X) \leq 1.0 \times 10^{-5} \quad 90\% \; \text{CL}
\ee
These limits are valid for  masses 13 MeV$< m_X < 80$ MeV.
A bound  weaker than the one from \eq{jodidio} is obtained for the opposite chirality.
Recently, a comparable limit has been provided in the region 87.0 MeV $< m_X <$ 95.1 MeV 
in \cite{Aguilar-Arevalo:2020ljq}.
Weaker limits are obtained for $m_X < 13$ eV, which includes the massless case.

The limit in \eq{twist} on the branching ratio
implies a  limit, valid for  $m_X\ll m_{\mu}$, on the scale of the left-handed ALP and the dark photon 
\be
\left|\frac{ f_a}{c_{\mu e}^{L}}\right|  \quad \mbox{and} \quad \left|\frac{\Lambda}{\sqrt{2} \,d_{\mu e}^{L}}\right|\geq 0.9 \times 10^{6} \;   \label{boundL}
\ee
again, in agreement with the analysis in \cite{Calibbi:2020jvd}. 

The limits above are among the strongest available  on the dark sector scale. They are stronger than those obtained from bounds on stellar cooling~\cite{Graham:2015ouw,Fabbrichesi:2020wbt} which are the strongest available constraints on the flavor conserving interactions.


The limits in the case of the massless dark photon requires some caution. If an UV completion is envisaged, the same diagram giving rise to the effective interaction in  $\mu^+\to e^+ X$ also enters in $\mu \to e \gamma$---which  brings in the very strong limit for that BR. While our discussion is at the level of the effective operators in \eq{L-DP}, if the underlying UV model needs to be addressed, we must assume a cancellation mechanism for the loop diagram with the photon to use the bounds in \eq{boundR} and \eq{boundL} . 

A possible mechanism  to suppress the new physics contribution to $\mu\to e \gamma$ decay amplitude induced at one-loop with respect to the amplitude for $\mu\to e \bar{\gamma}$, is the one proposed in \cite{Gabrielli:2019uqu}. It is based on an underlying global (dark) $SU(2)$ symmetry whose generators are commuting with the EM charge operator, but not with the dark charge operator associated to $U(1)_D$.
Due to the different group factors of the photon and dark-photon couplings of the new particles running in the loop, with respect to the dark $SU(2)$ group, the new physics contribution to the $\mu\to e \gamma$ amplitude at one-loop turns out to be vanishing with respect to the corresponding amplitude in $\mu\to e \bar{\gamma}$ \cite{Gabrielli:2019uqu}.

Concerning the constraints  $\mu\to e \gamma$ on the ALP contributions, this is not an issue since $\mu\to e \gamma$ is always loop suppressed with respect to $\mu\to e X$, for $X$ an ALP.

\vskip0.5cm
\textit{Distinguishing the spin.---} 
If the dark sector boson does couple off-diagonal flavor states, the $\mu^+ \to e^+ X$ decay is a most sensitive process and  one where a signal of the dark sector may first emerge. Yet the nature of such a state remains undetermined: it can be a spin 0, or a spin 1 or, even, a spin 2 particle. To find out in an efficient way which of these possibilities is realized, we need to measure the positron polarization.

The polarization of the positron  has been measured at  SIN \cite{Burkard:1985kf} for both polarized and unpolarized muons. More recently, both the longitudinal~\cite{Fetscher:2003ke,Prieels:2014paa,Fetscher:2007zz} and transverse~\cite{Danneberg:2005xv} positron polarizations  have been measured  in anti-muon decay. For our purposes, both the momentum and polarization of the positrons need to be measured as in the setup in~\cite{Prieels:2014paa}. The measure of the positron momentum comes from a spectrometer, that of the polarization relies on a polarimeter utilizing the spin dependence in the  Bhabha scattering and in-flight annihilation of the positrons. 

Assuming that a signal has been observed, how many events do we need in order to distinguish the spin of the dark sector boson?
The widths in \eqs{widthDP}{widthALP} show the dependence on the positron helicity.
Let us take fully polarized anti-muons with $P_\mu=1$ and consider the kinematical region where the SM background is suppressed. If the measured helicity of the positron is $\lambda_e=-1$, only the right-handed dark photon and ALP couplings contribute and they have clearly distinguishable  distribution in the variable $z\equiv \cos \theta$, namely $1+z$ and $1-z$.

The choice of the positron polarization $\lambda_e=-1$ is particularly favorable inasmuch as the SM background is strongly suppressed, as shown by Eq.(\ref{SM}).

The spin of the dark boson $X$ is determined by  comparing the distribution in $z$ of the collected events.  A likelihood analysis provides the means to find how many events we need in order to distinguish the spin with the desired confidence level (CL).

 \begin{figure}[t!]
\begin{center}
\includegraphics[width=3in]{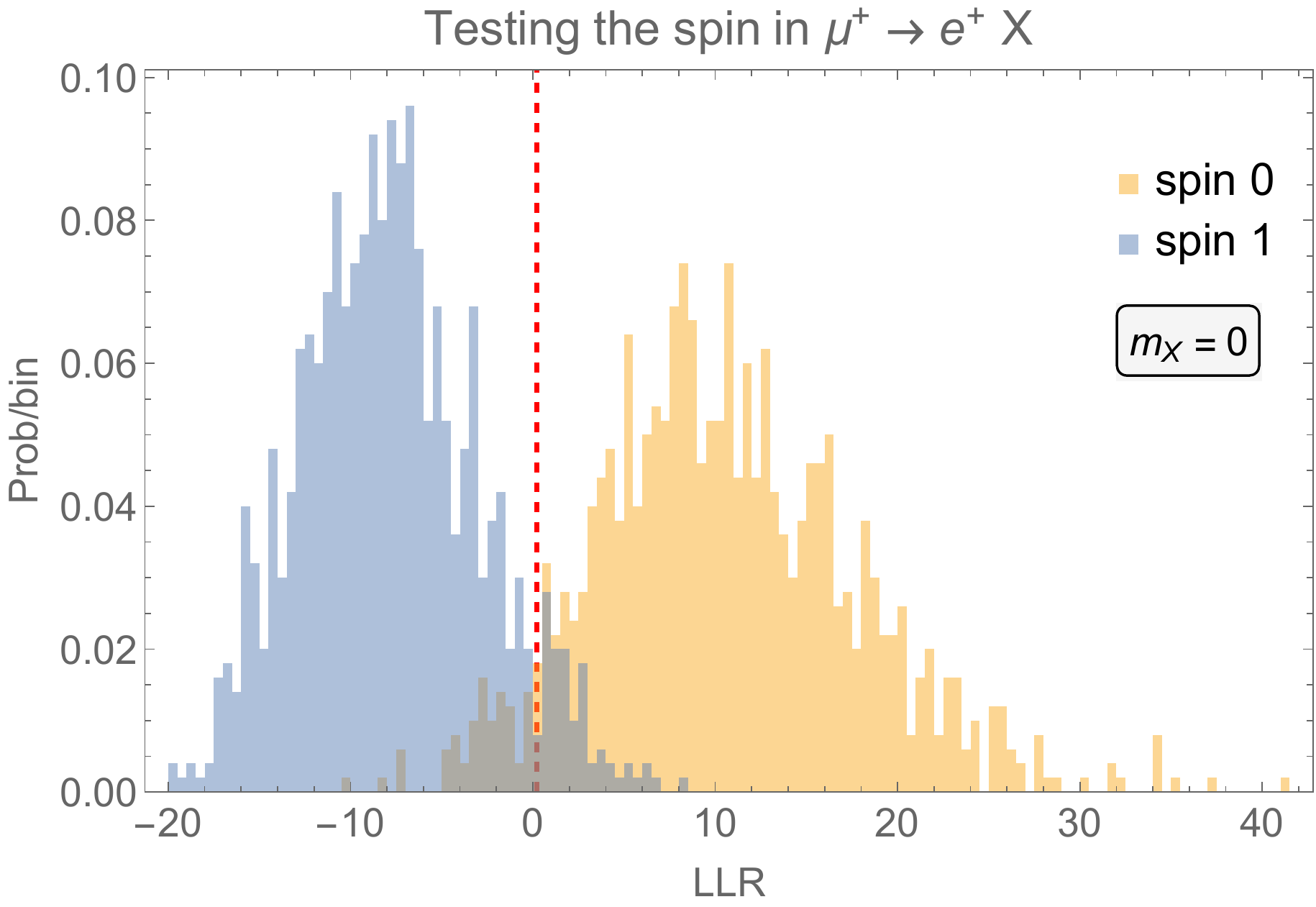}
\includegraphics[width=3in]{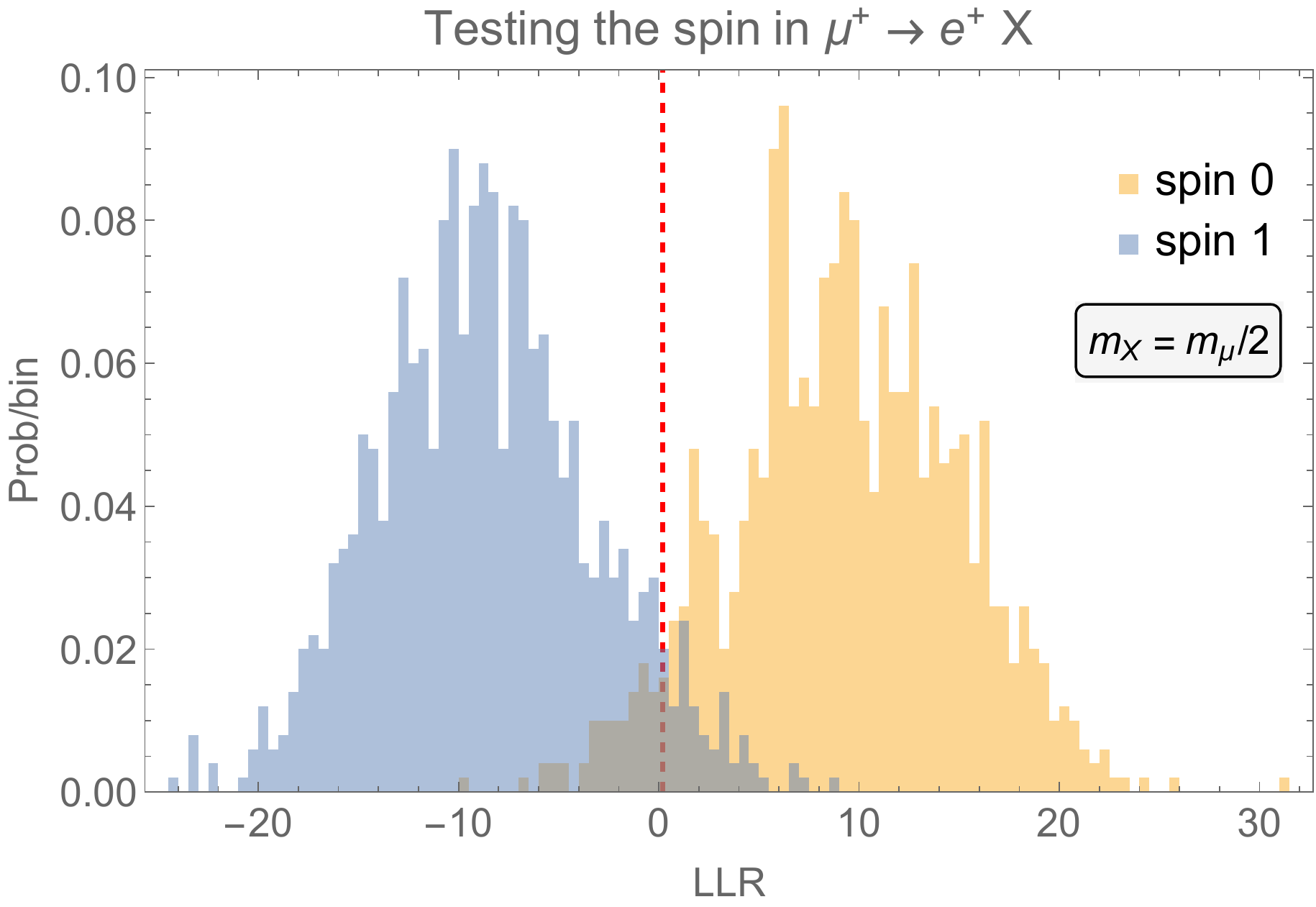}
\caption{\small Hypothesis test for spin 0 vs.\ spin 1 with 1000 pseudo-experiments. The positron helicity $\lambda_e=-1$ case.   \underline{Above}: Massless case: The histograms  correspond to  6 signal events and 20\% SM background. The test distinguishes the two spin hypothesis with a 99\% CL. \underline{Below}: Massive case for $m_X\simeq m_\mu/2$: The histograms correspond to  9 signal events and 20\% SM background. The test distinguishes the two spin hypothesis  with a 99\% CL. The red vertical line separates equal areas under the LLR curves.
\label{test} 
}
\end{center}
\end{figure}

The probability distribution functions (pdf) for the different spin hypotheses can be extracted directly by the dependence on $\cos \theta$ in \eqs{widthDP}{widthALP}.  They are well separated,  at least in the massless case.

Let us take $N_{\text{obs}}^{(J)}$ spin-$J$ signal events generated according to the corresponding pdf as well as $N_{\text{obs}}^{(\text{bkg})}$ background events. Each event $i$ is characterized by the value of  $z$. The  likelihood function for the spin  hypothesis $J^\prime$  is given by
\bea
{\cal L}_{\text{spin-}J^\prime}  &= & e^{-N_{\text{obs}}^{(J)}- N_{\text{obs}}^{(\text{bkg})}} \\
& &\!\!\!\!\! \!\!\!\!\!  \!\!\!\!\! \!\!\!\!\! \!\!\!\!\! 
\times \prod_i^{N_{\text{obs}}^{(J)}+N_{\text{obs}}^{(\text{bkg})} } \left[ N_{\text{obs}}^{(J)}  \times \text{pdf}_{J^\prime} (z_i)  +N_{\text{obs}}^{(\text{bkg})} 
\times \text{pdf}_{\text{bkg}} (z_i)  \nn
\right] \, ,
\eea
where the events $z_i$ are taken from the spin $J$ population. In this way, it is possible to randomly generate $N_{\text{obs}}$ events and compute the log of the likelihood ratio (LLR) defined by
\be
\text{LLR} =2\log \frac{{\cal L}_{\text{spin-0}} }{{\cal L}_{\text{spin-1}}}\, .
\ee
By repeating this pseudo-experiment $N_{\text{ps}}$ times, we construct a sample that can be used to compute the statistical distribution of the two spin hypotheses. We take $N_{\text{ps}}=10^3$

We construct two statistical samples for the LLR, the first one with events characterized by the value of  $z_i$ generated according to the spin-0 population, the second one with $z_i$ generated according to the spin-1 population.
In Fig.~\ref{test} we show our results for the LLR analysis. To quantify the difference in terms of statistical significance, we compute the value (indicated by the red vertical line in Fig.~\ref{test}) for which   the integral  under the curve for the spin 0 from $-\infty$ to that value is equal to the integral under the curve for the spin 1 from  that value to $\infty$. Let us call the value of these two identical integrals $p$.  The  significance is defined as
${\cal Z} = \Phi^{-1}(1-p)$
where
$\Phi(x) = \left[ 1 + \text{erf} \left( x/\sqrt{2} \right) \right]/2$.
The value of  ${\cal Z}$ assigns a statistical significance to the separation between the two LLR distributions. We can take $2 {\cal Z}$ as the number of $\sigma$'s, in the approximation in which the distribution is assumed to be Gaussian, and translates the number of $\sigma$'s into a CL.

 As shown in Fig.~\ref{test}, we reach a CL of 99\% with just $N=6$ events in the massless case, and $N=9$ for the massive one.
 We have assumed a conservative 20\% of background events due to the SM and taken their   angular distribution (in the region $x_e \simeq 1$) to be that of a spin 0 particle, as predicted by \eq{gamma}.

 This choice is suitable only for events having $\lambda_e=-1$ positron polarization---for which the SM background is strongly suppressed. Indeed, the residual background events for $\lambda_e=-1$, not shown in Eq.(\ref{SM}), come from corrections proportional to the positron mass and transverse polarizations, thus strongly suppressed.
 On the other hand, selecting events with opposite helicity $\lambda_e=1$---for which the leading SM contribution is dominant---a
larger percentage  of background should be assumed, since assuming 20\% background events would be too optimistic in this case.

\begin{figure}[t!]
\begin{center}
\includegraphics[width=3in]{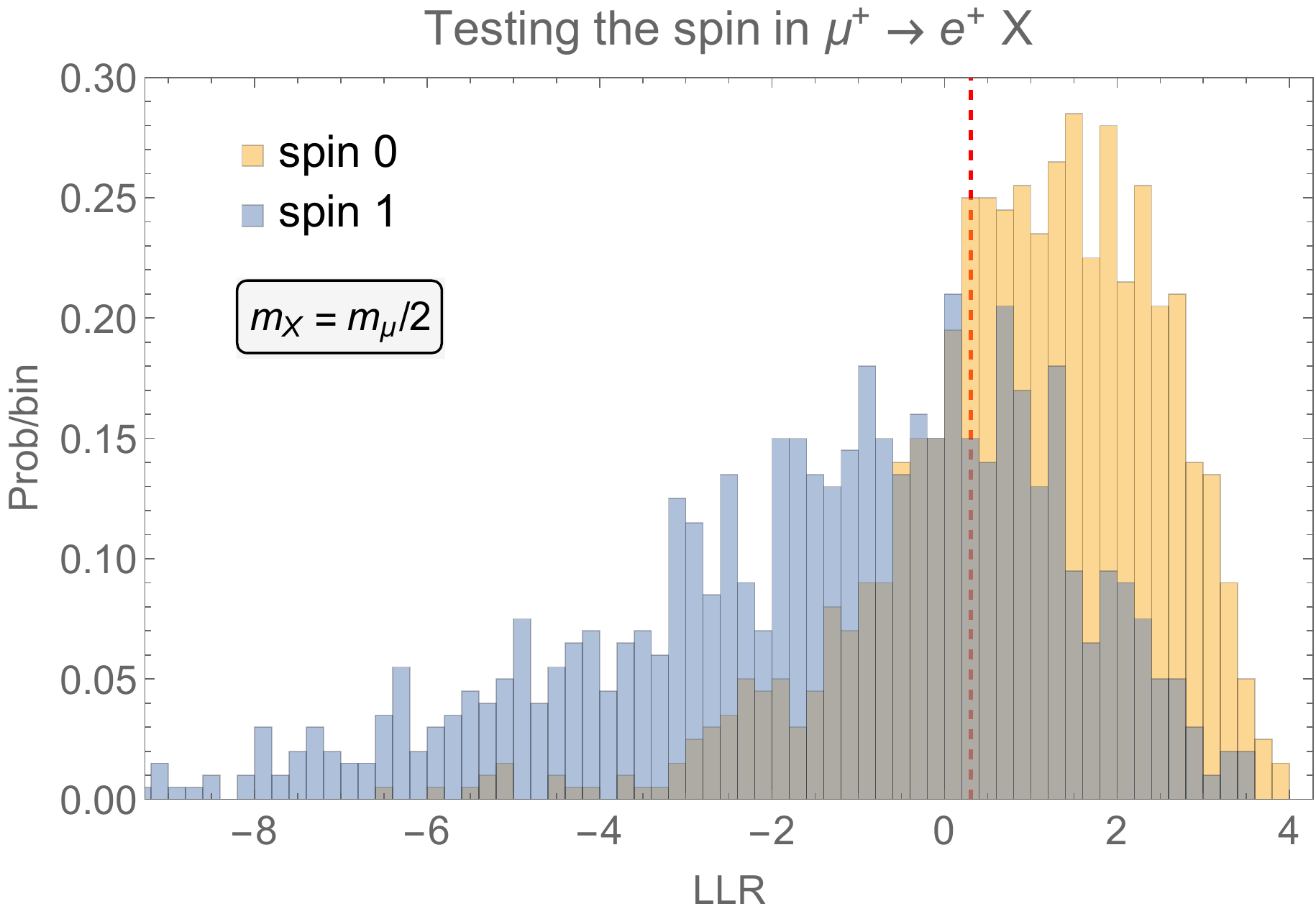}
\caption{\small Hypothesis test for spin 0 vs.\ spin 1 with 1000 pseudo-experiments without the use of the positron polarization for $m_X=m_\mu/2$. The histograms  correspond to  30 signal events and 20\% SM background. The test distinguishes the two spin hypotheses with  a 68\% CL. Many more events are required to reach a 99\% CL. The red vertical line separates equal areas under the LLR curves.
\label{test_nopol} 
}
\end{center}
\end{figure}

 The use of the positron helicity is much more efficient in separating the events than the small difference in the widths due to the mass correction. As an example, we run the same LLR using as pdf's those in \eqs{widthDP}{widthALP} after summing over the  positron polarizations and comparing the left-handed with the right-handed interactions. We take  $m_X = m_\mu/2$ and find (Fig.~\ref{test_nopol})  with $N=30$ events only 68\% CL in this case. Smaller values of $m_X$ leads to even more events being required in order to reach a given CL.
 
\vskip0.5cm
\textit{The spin 2 case.---}
Even though less plausible, we cannot exclude that the dark boson be 
a massive spin-2 fundamental particle, with graviton-like couplings, that is universally coupled to the energy-momentum tensor of all SM fields.

Fundamental massive spin-2 fields have been predicted by several extensions of gravity theories, like the massive Kaluza-Klein  excitations of the massless graviton in theories with large extra-dimensions~\cite{ArkaniHamed:1998rs,Randall:1999ee}  as well as the massive graviton in the bi-metric theories \cite{Schmidt-May:2015vnx}.

We stress  that, although we consider a spin-2 particle universally coupled to the energy-momentum tensor of all SM fields as for massive gravitons, this scenario has nothing to do with massive quantum gravity theory, since our spin-2 field (that could be also a composite particle) does not include tree-level self interactions as predicted by massive quantum gravity theories. We have introduced this scenario for merely  phenomenological purposes, to just account for a massive $X$ spin-2 particle in the angular distributions of the $\mu \to e X$ process.
Moreover, the request of universal coupling to all SM fields guarantees, by means of Ward identities, that no $1/m_X$ mass term singularities appear in the scattering processes with external spin-2 fields, extending the validity of the effective theory up to energies of the order of the
scale of the spin-2 $\Lambda_{\rm S2}$ coupling.

Although we do not make any particular assumption on the spin-2 mass $m_X$, one should bear in mind that its value  is constrained to  the mass range $1 {\rm eV}\lsim  m_X \,\lsim \,m_\mu$
in order to avoid the strong constraints from searches on the Newton law deviations at short distance \cite{Murata:2014nra} (lower bound) and the requirement of decaying outside the detector (upper bound). For masses in this range, a viable scale $\Lambda_{\rm S2}$ for the spin-2 can be $O(1)$ TeV~\cite{Comelato:2020cyk}.

As shown in \cite{Degrassi:2008mw}, as a consequence of Ward identities, a spin-2 field effectively coupled to matter fields at tree-level with universal coupling
can develop finite contributions to the 
 flavor changing couplings, via one-loop electroweak corrections to the quark energy-momentum tensor. Following the results of \cite{Degrassi:2008mw}, the amplitude at low energy for the   transitions $\mu \to e X$, where $X$ is now a spin 2 graviton-like particle,  can be then parametrized as
\bea
M(\mu \to e X)&=&\frac{1}{\Lambda^3_{\rm S2}}\Big\{g^L_{\mu e}\Big[\bar{e} (p_2) V_L^{\mu\nu}(p,q)
  \mu (p_1)\Big] \nn  \\ & & \!\!\!\!\!\!  \!\!\!\!\!\! \!\!\!\!\!\!  \!\!\!\!\!\!  + \;
 g^R_{\mu e}\Big[\bar{e}^j(p_2) V_R^{\mu\nu}(p,q)
  \mu^i(p_1)\Big]\Big\}\epsilon_{\mu\nu}(q)\, ,
\label{ampG}
\eea
where $q=p_1-p_2$, $p=p_1+p_2$, and $\epsilon_{\mu\nu}(q)$ is the polarization vector of the massive spin-2 particle of momentum $q$;
  $g^{L,R}_{\mu e}$ are some coefficients that arise at 1-loop. In \eq{ampG} we have reabsorbed the mass-scales $\Lambda^{L,R}_{\mu e}$  arising from the 1-loop computation of the off-diagonal matrix element of energy-momentum tensor  (which are independent by $\Lambda_{\rm S2}$) into the definition of the dimensionless couplings, that modulo a sign are $g^{L,R}_{\mu e}\equiv \Lambda^2_{\rm S2}/(\Lambda^{L,R}_{\mu e})^2$. The tensors $V_{L,R}^{\mu\nu}(p,q)$ for the on-shell fields are given by
\bea
V^{\mu\nu}_{L,R}(p,q)=\left(\gamma^{\mu}p^{\nu}+\gamma^{\nu}p^{\mu}\right)m_G^2 \frac{(1\pm \gamma^5)}{2}\, .
\label{VG}
\eea

Due to the angular momentum conservation, the amplitude in \eq{ampG}  
with a strictly massless spin-2 particle like the Einstein graviton ($m_X=0$) is vanishing. On the other hand, in the case of a massive spin-2, a non-vanishing limit $m_X\to0$ survives due to the presence of spin-2 longitudinal polarizations.

The differential  decay width for the massive spin 2 particle (S2) is given by
\begin{widetext}
\bea
\left. \frac{\di \Gamma}{\di\cos \theta}\right| _{\text{\rm S2}} &=& \frac{m_\mu^7}{\Lambda_{\rm S2}^6} \frac{(1 - r)^4}{48 \pi} \Big\{ |g_{\mu e}^R|^2(1- \lambda_e)  \left[ (1- P_\mu   \cos \theta ) + \frac{3}{2}r ( 1 + P_\mu   \cos \theta) \right]  \Big.\nn  \\
& & +  \left. |g_{\mu e}^L|^2 (1+ \lambda_e)  \left[ 1+  P_\mu   \cos \theta  + \frac{3}{2}r( 1 - P_\mu   \cos \theta) \right]  \right\} \label{2} \, ,
\eea
\end{widetext}
corresponding, for $X$ massless,  to an unpolarized ${\rm BR(\mu^+ \to e^+ X)} = 2.6\times 10^{-8} ({\rm TeV}/\Lambda_{\rm S2})^6$ for \hbox{$g_{\mu e}^L=g_{\mu e}^R=1$}.

The differential width in \eq{2}, for positron helicity $\lambda_e=-1$, has the same angular dependence as in the ALP case at fixed positron polarization, in the limit of vanishing dark boson mass.

\begin{figure}[t!]
\begin{center}
\includegraphics[width=3in]{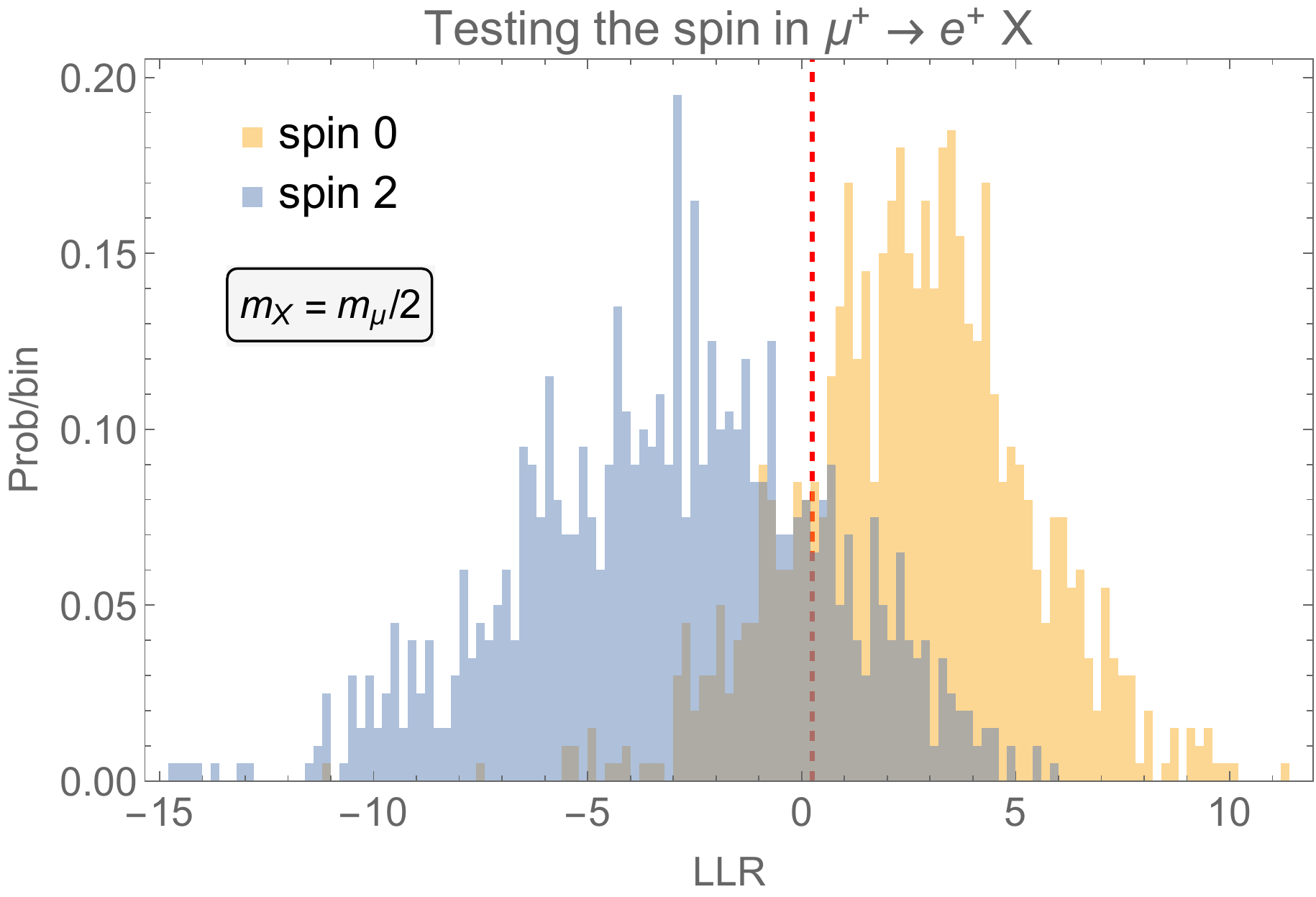}
\caption{\small Hypothesis test for spin 2 vs.\ spin 0 with 1000 pseudo-experiments.  The positron helicity $\lambda_e=-1$ case for $m_X=m_\mu/2$.   The histograms  correspond to  30 signal events and 20\% SM background. The test distinguishes the two spin hypothesis  with a 95\% CL. The red vertical line separates equal areas under the LLR curves.
\label{testG} 
}
\end{center}
\end{figure}

Measuring the positron polarization could be very efficient in distinguishing spin-1 versus spin-2, due to the different angular dependence in the $m_X$ independent part, as in the spin-0 versus the spin-1.
Moreover, it can still be  distinguished from the spin-0 and spin-2, but only for sizable values of the mass $m_X$.
In this case, we find (see Fig.~\ref{testG}) that at least $N=30$ events are required (for $m_X=m_\mu/2$) to reach a 95\% CL.

\vskip2em
\textit{The $\mu^-  \to e^- $ conversion in muonic atoms.---} Muon conversion $\mu^-  \to e^- X$ is  another promising process where to look for the dark sector. The results of the SINDRUM II collaboration~\cite{Bertl:2006up} can be used to set a limit~\cite{GarciaiTormo:2011et} $\BR(\mu\to e X) < 3 \times 10^{-3}$ and ongoing experiments of the  Mu2e \cite{Bartoszek:2014mya} and COMET \cite{Adamov:2018vin} collaborations will push this limit  further.  Assuming a discovery,  the electron helicity could be used to distinguish  the nature of the $X$ boson in an efficient manner---which can be compared to  an analysis~\cite{Uesaka:2020okd} based only on the differences in the energy spectrum---but \eqs{widthDP}{widthALP} need to be adapted  to take into account the wave-functions of the bounded muons. 

\vskip2em
\textit{The $\tau\to \mu  X$ and $\tau\to  e  X$ decays.---} The same expressions in \eqs{widthDP}{widthALP}  apply, \textit{mutatis mutandis}, in the case of the decays of the $\tau$ into  a muon or an electron. Limits  for these processes have been set by the ARGUS collaboration ~\cite{Albrecht:1995ht} and Belle~\cite{Yoshinobu:2017jti} to branching ratios of the order of $O(10^{-3})$. Even though these bounds are less stringent than those for the anti-muon decay, they are expected (see \cite{Calibbi:2020jvd}) to be improved by two ($\tau\to \mu  X$) and three ($\tau\to  e  X$) orders of magnitude. Though the same argument for using the helicity of the final lepton  to reduce the SM background and distinguish the spin of the particle $X$ might be applied also to these decays, 
 the problem of the full reconstruction of the $\tau$ boost necessary  to define the $\tau$ rest frame makes the analysis  more  complicated.

\vskip2em
\textit{Conclusions.---} The $\mu^+ \to e^+ X$ decay, where $X$ is a dark sector boson, provides one of the most stringent  
bounds on the scale of dark sector interactions. We have shown that---in the presence of a signal---the simultaneous measurement of the positron polarization and angular distribution of the photon momentum makes possible  to disentangle the nature of the dark sector portal in the case of the spin-0 ALP  and the massless spin-1 dark photon. Instead, the  massive  spin-1, dimension 4 dark portal  remains undistinguishable from the axion-like case.
The simultaneous measurement of the positron polarization and angular distribution of photon  is also useful for controlling  the SM background.

\vskip2em
\textit{Acknowledgments.---}
       {\small
MF is affiliated to the Physics Department of the University of Trieste and the Scuola Internazionale Superiore di Studi Avanzati---the support of which is acknowledged. 
MF and EG are affiliated to the Institute for Fundamental Physics of the Universe, Trieste, Italy.}



\end{document}